\documentclass[twocolumn,showpacs,preprintnumbers]{revtex4}

\usepackage{graphicx}
\usepackage{dcolumn}
\usepackage{bm}

\begin{document}
\title{Raman Spectra of Triplet Superconductor in Sr$_2$RuO$_4$}
\author{Hae-Young Kee$^{1}$, K. Maki$^{2}$, and C. H. Chung$^{1}$}
\affiliation{$^1$ Department of Physics, University of Toronto, Toronto,
Ontario M5S 1A7, Canada\\
$^2$ Department of Physics, University of Southern California, Los Angeles, CA 90089}
\date{\today}

\begin{abstract}
We study the Raman spectra of spin-triplet superconductors in Sr$_2$RuO$_4$.
The p-wave and f-wave symmetries are considered.
We show that there is the clapping mode with frequency
of $\sqrt{2} \Delta(T)$ and $1.02 \Delta(T)$ for
p-wave and f-wave superconductors, respectively.
This mode is visible as a huge resonance in the B1g and B2g modes 
of Raman spectra.
We discuss the details of the
Raman spectra in these superconducting states.

\end{abstract}         

\maketitle

\def\be{\begin{equation}}
\def\ee{\end{equation}}
\def\bea{\begin{eqnarray}}
\def\eea{\end{eqnarray}}
 
\section{Introduction}
The superconductivity in Sr$_2$RuO$_4$ was discovered in 1994.\cite{maeno}
Shortly after the discovery of superconductivity, 
a possible triplet p-wave superconductivity
with the following order parameter was postulated 
by Rice and Sigrist.\cite{rice, maeno2}
\begin{equation}
{\hat \Delta}({\bf k})=\Delta {\hat d} (k_x \pm i k_y),
\label{p-wave}
\end{equation}
where
$\Delta$ is the magnitude of the superconducting order parameter.
Here ${\hat d}$-vector is called spin vector perpendicular to
the direction of the the spin associated with the condensed pair.\cite{vollhardt}
Notice that this state is analogous to the $A$ phase of $^3$He
and there is a full gap on the Fermi surface.
Subsequently, the triplet superconducting nature has been 
confirmed by the constancy of $^{17}$O Knight shift
(spin susceptibility) through $T_c$ for the magnetic field parallel
to the $a-b$ plane\cite{ishida}. 
The broken time reversal symmetry state have been also confirmed by
the spontaneous magnetic moment found in $\mu$SR experiment\cite{luke}. 
In general, the triplet superconductors have a variety of 
collective modes.  The spin waves and the clapping mode
with the order parameter of Eq. \ref{p-wave} were studied
in Ref. \cite{tewordt,kee,kee2}.
Further, the effect of the clapping mode on the sound wave 
was studied.\cite{kee2}  However, this coupling  is very small
to detect the existence of the clapping mode.

In the meantime, the sample quality of Sr$_2$RuO$_4$ has been improved.
The cleanest sample shows the transition temperature close
to the optimal $T_c =1.5K$ deduced from the $T_c$
dependence of the residual resistivity.\cite{mackenzie}
All these high quality samples exhibit the character of
nodal superconductors;
the $T^2$ behavior of the specific heat\cite{nishizaki}, the $T$-linear dependence
of superfluid density\cite{bonalde}, the $T^3$ behavior
of $1/T_1$ in NMR\cite{ishida3}, the $T^2$ behavior
of the ultrasonic attenuation\cite{lupien},
and the $\sqrt{H} $ dependence of
the specific heat in a magnetic field\cite{nishizaki}
at low temperature. 
These low temperature behaviors of the specific heat and superfluid
density are consistent with a f-wave superconductor with nodes
in the order parameter.\cite{hasegawa,won,dahm,miyake,graf}

On the other hand, the quasi-two dimensional system with strong paramagnon
(the ferromagnetic spin fluctuation) favors the p-wave superconductor
as given in Eq. \ref{p-wave}.\cite{sigrist, sato}
In order to understand this situation, the nodal structure of
the order parameter becomes of crucial importance.
In a series of papers, Won and Maki have shown that the magneto-thermal
conductivity for $T \ll {\tilde v} \sqrt{e H} \ll \Delta (0)$
will provide the direct access to this question\cite{won2,won3},
where ${\tilde v}=\sqrt{v_a v_c}$ and $v_{a,c}$ are the anisotropic Fermi velocities.
Such experiments have been carried out by Tanatar {\it et al}\cite{tanatar} and
Izawa {\it et al}\cite{izawa}. 
The result is consistent with the f-wave superconductor with
horizontal nodes of Eq. \ref{f-wave}, even though the thermal conductivity
data cannot exclude a small admixture, a few \% of p-wave order parameter, 
Eq. \ref{p-wave}.
\begin{equation}
{\hat \Delta }({\bf k}) = \Delta {\hat d}[ (k_x \pm i k_y )\cos{(c k_z)} ],
\label{f-wave}
\end{equation}
where $c$ is the lattice constant along the $c$-axis.

Parallel to this development, Zhitomirsky and Rice\cite{zhito}
have proposed an alternative model, multi-gap model for Sr$_2$RuO$_4$.
As it is known that the Fermi surfaces in Sr$_2$RuO$_4$ consist
of three different bands labeled by $\alpha$-, $\beta$-, and
$\gamma$-bands\cite{maeno2}, it is also believed that the superconductivity
arises mainly in the $\gamma$-band.
It was proposed that a full gap with p-wave, Eq. (1) exists
in the active band $\gamma$, while line nodes with f-wave order parameter, Eq.
\ref{f-wave-zr}
develops in the $\alpha$ and $\beta$ due to proximity effect.\cite{zhito}
\begin{equation}
{\hat \Delta }_{zr} ({\bf k}) = \Delta {\hat d}[ (k_x \pm i k_y )\cos{(c k_z/2)} ].
\label{f-wave-zr}
\end{equation}
While this model could reproduce the specific heat data by Nishizaki
et al\cite{nishizaki}, and the magnetic penetration depth data by
Bonalde et al\cite{bonalde,kusunose},  they have not attempted to calculate
the magneto-thermal conductivity which should be more revealing.
Note that the f-wave superconductor associated with $\alpha$ and
$\beta$-bands in Ref. \cite{zhito} is similar to the Eq. \ref{f-wave},
but not the same. 
The model with ${\hat \Delta}_{zr}$ produced much larger ($\sim$ 30 times)
$cos{(2\phi)}$ term in the angular dependence of magneto-thermal
conductivity.
The better test of the nodal position can be done through the angle (azimuthal
angle of ${\bf k}$ from the $a$-axis)
dependence of the thermal conductivity\cite{dahm,maki}.

More recently, Deguchi et al observed a double transition in the specific heat with
the magnetic field near $H_{c2}$, and  
this experimental result was interpreted in terms of multigap model.\cite{deguchi}
On the other hand, the behavior of the specific heat 
and the magnetic penetration depth for
low temperature $(T \ll T_c)$  and the low field $(H \ll H_{c2})$
 appears to be consistent
with the f-wave order parameter given in Eq. \ref{f-wave}.
At the moment, it is not clear whether the order parameter of Eq. \ref{f-wave},
or the multigap model with the order parameters of Eq. \ref{p-wave}
and Eq. \ref{f-wave-zr} is adequate.
This is the fundamental issue for Sr$_2$RuO$_4$.

In this paper, we study the Raman spectra of spin triplet superconductors
in Sr$_2$RuO$_4$. 
We consider three different order parameters given in Eq. \ref{p-wave},
\ref{f-wave}, and \ref{f-wave-zr} sketched in Fig. 1;
(a) p-wave with a full gap, (b) f-wave with horizontal nodes, and
(c) f-wave with node at the zone boundary.
We identify the clapping mode\cite{vollhardt} in these superconductors and
consider its effect on Raman spectra.
We show that (1) the Raman spectroscopy can detect the clapping mode,
and (2) it can discriminate the multigap model from the single gap model.
\begin{figure}
\scalebox{.52}{\includegraphics{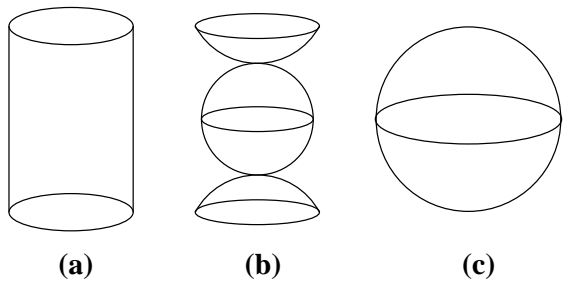} }
\caption{Order parameters with the cylindrical Fermi surface for 
(a) p-wave (b) f-wave of the Eq. \ref{f-wave}  
(c) f-wave of the Eq. \ref{f-wave-zr}  }
\label{fig1}
\end{figure}
For the p-wave superconductor, the frequency of the clapping mode 
was found at  $\sqrt{2} \Delta (T)$\cite{kee}, and 
this mode exists as a sharp resonance in $B_{1g}$ and $B_{2g}$ modes
of Raman spectra. 
For the f-wave superconductors (Eqs. \ref{f-wave}
and \ref{f-wave-zr}), the frequency of the clapping mode
gets smaller than that of the p-wave, $1.02 \Delta$ as expected
due to the existence of node, and this mode is also detectable
in $B_{1g}$ and $B_{2g}$ modes of Raman spectra.
While the weighted sum of the clapping mode contributions
from the p-wave and the f-wave appears in the multigap model, 
the single gap model of Eq. \ref{f-wave}
has the contribution from solely the clapping mode in the f-wave
superconductor.

The paper is organized as follows.
The formalism to identify the clapping mode and 
its effect on the Raman spectra is summarized in the section II.
The results of the Raman spectra of $B_{1g}$ and $B_{2g}$ modes
for p-wave and f-wave superconductors
are presented in the section III and IV, respectively.
The conclusion and discussion will be followed in the section V.
The details of computing the correlation functions are presented
in Appendix.

\section{Formalism of Raman spectra and clapping mode}
The electronic Raman scattering in superconductors are well described
in Ref. \cite{devereaux}. Therefore, here we give a brief summary
of the Raman spectra. 
The Raman spectra $S_i$ in superconductors measure
effective density fluctuations, and its strength of the scattering
is determined by the Raman vertex, $\gamma_i$.
One can select the Raman vertex which allows for different projections
on the Fermi surface. The intensity of the each Raman mode
provides information on the gap structure along the Fermi surface.
The Raman spectra in superconductors is determined from
\begin{equation}
S^{0}_i (\omega, {\bf q} \rightarrow 0)=
Im \left[ \langle \gamma_i, \gamma_i \rangle -
\frac{ \langle \gamma_i, 1 \rangle}{\langle 1,1 \rangle} \right].
\label{raman_S}
\end{equation}
Here we use the following notational convenience,
\begin{equation}
\langle A, B \rangle =
T \sum_n \sum_{\bf p} {\rm Tr} [ A
\rho_3 G({\bf p},\omega_n)
B \rho_3 G({\bf p}-{\bf q},i\omega_n-i\omega_{\nu})] \ ,
\end{equation}
where the single particle Green's function, $G (i\omega_n,{\bf k})$,
in the Nambu space is given by
\begin{equation}
G^{-1}(i\omega_n,{\bf k})=i \omega_n - \xi_{\bf k}\rho_3
- \Delta ({\hat k}\cdot {\hat \rho}) \sigma_1.
\label{green}
\end{equation}
Here $\rho_i$ and $\sigma_i$ are Pauli matrices acting on 
the particle-hole and spin space, respectively.
$\omega_n = (2n+1) \pi T$ is the fermionic Matsubara
frequency, and $\xi_{\bf k}=(k_x^2+k_y^2)/2m - \mu$ where
$\mu$ is the chemical potential.
The Raman vertics, $B_{1g}$ and $B_{2g}$ are written as 
\begin{eqnarray}
\gamma_{B1g} &=& \sqrt{2} \cos{(2\phi)} 
\nonumber\\
\gamma_{B2g} &=& \sqrt{2} \sin{(2\phi)},
\end{eqnarray}
where $\phi$ is the angle of the wave-vector ${\bf k}$ on the Fermi surface.
The second term of Eq. \ref{raman_S} is a back-flow term due to
the charge conservation, which can be seen in the limit
of $\gamma_i =$ const. In this case, Raman intensity should vanishes
because there is no density fluctuation in the homogeneous limit of
${\bf q} \rightarrow 0$ in a superconductor.

The clapping mode is the fluctuation of the order parameter
which can be written as $\delta \Delta \rho_3
\sim \exp{ (\pm 2 i \phi) } (\sigma_1 \pm i \sigma_2) \rho_3$ in the Nambu space,
As it is indicated in the fluctuation of the order parameter, $\delta \Delta
\propto \exp{(\pm 2 i \phi) }$, 
this mode can directly couple to $B_{1g}$ and $B_{2g}$ channels
of the Raman spectra.
The clapping mode makes the additional contribution to
the Raman intensity given by
\begin{equation}
S_i = S^{0}_i +
{\rm Im} \left( \frac{\langle \gamma_i ,\delta \Delta  \rangle
\langle \delta\Delta, \gamma_i  \rangle}
{g^{-1} - \langle \delta \Delta ,\delta \Delta \rangle} \right) \ ,
\label{total_S}
\end{equation}
where $g$ is the coupling constant which mediates superconducting state,
and the coupling between the fluctuation of the order parameter
and the light scattering with Raman vertex, $\gamma_i$.
On the other hand, the $A_{1g}$ mode does not couple to the clapping mode.
Therefore $S_{A_{1g}}$ is given by the first term in Eq. \ref{total_S}. 
Note that $S^0_i$ is the same for all three modes ($B_{1g}$, $B_{2g}$,
and $A_{1g}$) for each order parameter.
This indicates the existence of the axial symmetry and horizontal nodes 
in the order parameter, if there is any.

\section{Raman spectra of $B_{1g}$ and $B_{2g}$ in p-wave superconductor}
The order parameter of Eq. \ref{p-wave} has a full gap on
the Fermi surface, therefore the bare Raman intensity in p-wave superconductor
is same as that of s-wave superconductor.
\begin{equation}
S^{0}_{B_{1g}} = Im\langle \cos{(2\phi)}, \cos{(2\phi)} \rangle
= \frac{2 \pi N(0) \Delta^2}{\omega \sqrt{\omega^2-4\Delta^2}}
\theta{(\omega^2-4\Delta^2)}.
\end{equation}
Note that the Raman intensity is zero for the frequency, $\omega < 2 \Delta$
due to the presence of the full gap.

The coupling to the collective mode leads to the additional
contribution to the Raman spectra.
While the $A_{1g}$ mode has the same frequency dependences as $B_{1g}$ and
$B_{2g}$ as far as the bare Raman intensity, $S^0$ is concerned,
this mode does not couple to the clapping mode.
The Raman intensity due to the clapping mode can be obtained 
by computing the correlation functions presented in the Appendix.
Here we summarize the result of the correlation functions.
\begin{eqnarray}
Re \langle \delta \Delta, \gamma_{B_{1g}} \rangle
&=& N(0) \left[ \omega \Delta f \right],
\nonumber\\
Re \langle \delta\Delta, \delta \Delta \rangle
&=& g^{-1} - N(0) \left[ \left( \frac{\omega^2}{2}-\Delta^2 
\right) f \right] \ ,
\end{eqnarray}
where $f$ is given by
\begin{equation}
f (\omega,T) =  \int_{\Delta}^{\infty}
dE \frac{\tanh{(E/2T)}}{\sqrt{E^2-\Delta^2}}
\frac{1}{4 E^2-\omega^2} \ .
\label{super}
\end{equation}
Since the imaginary part of 
$ \langle \delta \Delta, {\rm cos}(2\phi) \rangle $ and
$ \langle \delta\Delta, \delta \Delta \rangle $ are zero
for the frequency, $\omega < 2\Delta$,
the clapping mode appears as a resonance in the Raman spectra
for $B_{1g}$, which is given by
\begin{eqnarray}
S_{B_{1g}}^{clapp}(\omega) &= &
Im \left(  \frac{\langle \gamma_{B_{1g}},\delta \Delta  \rangle
\langle \delta\Delta ,\gamma_{B_{1g}} \rangle}
{g^{-1} - \langle \delta \Delta ,\delta \Delta \rangle }
\right) \nonumber\\
&=& 2\pi N(0) \omega^2 \Delta^2
f(\omega,T)\; \delta{(\omega^2 - 2\Delta^2)} .
\end{eqnarray}
The Raman intensity, $B_{1g} (\omega)$ is plotted in Fig. 2 with 
a finite impurity scattering, $\Gamma = 0.1 \Delta$.
It is important to note that there is a resonance at
the frequency of $\omega = \sqrt{2} \Delta$ with a Lorentzian shape.
The same is true for $B_{2g}$ channel.
\begin{figure}
\scalebox{.50}{\includegraphics{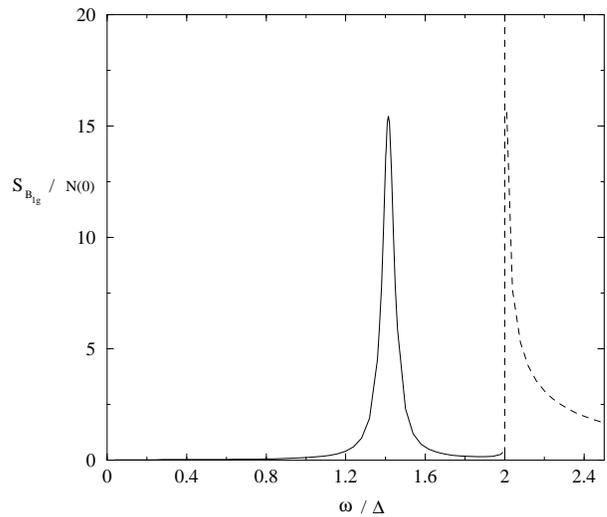} }
\caption{Raman intensity for p-wave with a finite impurity scattering,
$\Gamma= 0.1 \Delta$.
The solid line is the contribution from the clapping mode,
the dotted line is the bare Raman intensity}
\label{fig2}
\end{figure}

\section{Raman spectra of $B_{1g}$ and $B_{2g}$ in f-wave superconductor}
We consider the f-wave order parameter of Eq. \ref{f-wave}, and
the cylindrical Fermi surface independent of $k_z$.
Since there is a node along the $k_z$ direction, we expect 
the finite bare Raman intensity, $\langle\gamma_i,\gamma_i\rangle$
at low frequency,
\begin{eqnarray}
S^{0}_{B_{1g}}
& =& Im \langle \cos{(2\phi)}, \cos{(2\phi)}  \rangle  
\nonumber\\
& & \hspace{-1.5cm} =
\frac{ 4 N(0) \Delta }{\omega}
\left[ K(\frac{\omega}{2\Delta}) - E(\frac{\omega}{2\Delta})\right] 
\tanh{\frac{\omega}{4T}},
\end{eqnarray}
where $K$ and $E$ are complete Elliptical integral of first and
second kinds, respectively.

The contribution from the coupling to the clapping mode is given by
\begin{equation}
S^{clapp}_{B_{1g}}
 =  N(0) \omega^2 \Delta^2 
\frac{ 4 \omega_c^R \omega_c^I D(\omega,T)}
{(\omega^2-\omega_C^R)^2 + 4 (\omega_c^R \omega_c^I)^2 },
\end{equation}
where 
\begin{equation}
 D(\omega,T) = \frac{2}{\pi} 
 \int_{0}^{\infty} \frac{dE}{4 E^2 - \omega^2}
\frac{1}{\Delta}
\left( K(\frac{E}{\Delta})- E(\frac{E}{\Delta}) \right) 
\tanh{\frac{E}{2T} } 
\label{f_super}
\end{equation}
The frequency of the clapping mode and its damping are obtained as 
\begin{eqnarray}
\omega_c^R  &=& 1.02 \Delta
\nonumber\\
\omega_c^I & =&  0.57 \Delta,
\end{eqnarray}
where 
the $\omega_c^I$ is computed from the formula,
$\frac{{\rm Im}\langle \delta \Delta, \delta \Delta \rangle}
{ N(0) \omega_c^R D(\omega,T)}$, and 
${\rm Im} \langle \delta \Delta, \delta \Delta \rangle $ is given in
the Appendix, Eq. \ref{Im_f}.
The Raman intensity due the clapping mode is shown in Fig. 3.
It is important to note that the shape of the Raman intensity is
not Lorentzian, but it has asymmetry.
The Raman spectra for the another proposed f-wave, Eq \ref{f-wave-zr} 
are identical to the result presented here for the order parameter
of Eq. \ref{f-wave}. 
\begin{figure}
\scalebox{.50}{\includegraphics{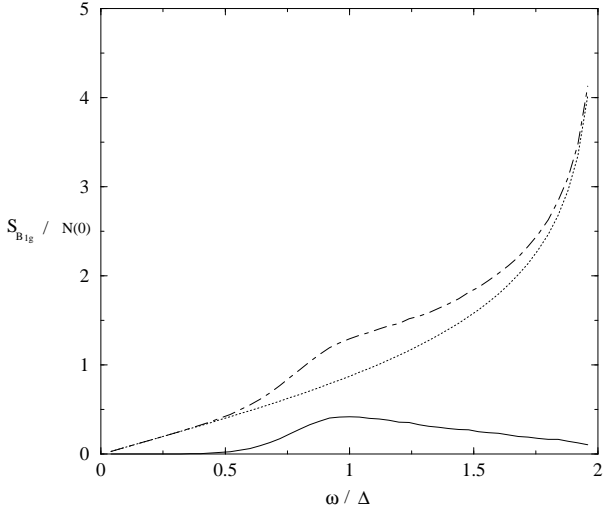} }
\caption{Raman intensity for f-wave. 
The solid line is the contribution from the clapping mode,
the dotted line is the bare Raman intensity, and the dashed line
is the total Raman intensity}
\label{fig3}
\end{figure}

\section{Conclusion}
Within the framework of the weak-coupling BCS theory, we have studied
the Raman spectra of two-dimensional p-wave (Eq. \ref{p-wave}) and 
f-wave (Eqs.  \ref{f-wave} and \ref{f-wave-zr}) 
superconductors; the candidate for Sr$_2$RuO$_4$.
The clapping mode with angular momentum $\pm 2$, parallel to
the c-axis is common to these ground states, which break the chirial symmetry.
We have shown that the clapping mode has the frequency of $\sqrt{2}\Delta(T)$,
and $1.02\Delta(T)$ for p-wave and f-wave superconductors, respectively.
We have also shown that the clapping mode couples both sound wave\cite{kee},
and $ B_{1g}$ and $B_{2g}$ modes of the Raman spectra.
While the coupling to the sound wave is very small, the clapping mode
appears as a huge resonance in Raman spectra.
Therefore, if the present experimental difficulty is overcome,
the Raman spectroscopy provides a unique window to probe the clapping mode.

Investigating the clapping mode in Raman spectra
will also probe the order parameter of Sr$_2$RuO$_4$.
The single model of the f-wave 
shows only contribution from the clapping mode of f-wave order parameter( Fig. 3 ).
On the other hand, 
the multigap model proposed by Zhitomirsky and Rice should be the weighted
sum of the p-wave (Fig. 2) and f-wave (Fig. 3) contribution.
It is important to note that the maximum gap of 
$\Delta_{zr} \approx (0.2 \sim 0.5) \Delta_{BCS}$ on the $\alpha$- and $\beta$-bands,
while the maximum of $\Delta$ on $\gamma$-band with p-wave 
is set to be $\Delta_{BCS}$ which is the energy scale we used in our figures.
Therefore, for the multigap model, the contribution from f-wave order parameter,
Eq. \ref{f-wave-zr} should be peaked around 
$(0.2 \sim 0.5) \Delta$ (with the contribution
from p-wave around $\sqrt{2} \Delta$), while
single gap model gives the peak around $ 1.02  \Delta$.
The Raman spectroscopy can discriminate the multigap model
from the single gap model.

{\it Acknowledgments}
This work was supported  by Canadian Institute for
Advanced Research, Canada Research Chair (H.-Y. K), 
and Natural Sciences and Engineering
Research Council of Canada. (H.-Y. K, C.-H. C)
K. Maki would like to thank Y. Matsuda for a useful discussion.

\appendix
\section{}
\subsection{the clapping mode and its contribution on Raman, $B_{1g,2g}$ in p-wave
superconductor}
The fluctuation of the order parameter corresponds to the
clapping mode, and it can be written as
$\delta \Delta \rho_3 \sim e^{\pm 2i\phi} (\sigma_1\pm i\sigma_2) \rho_3$.
Its coupling to the $\gamma_{B1g}$ mode of Raman spectra,
and its propagator, $\langle \delta\Delta, \delta\Delta \rangle$ 
are obtained by computing the following correlation function
within weak coupling theory.
\begin{eqnarray}
\langle \delta\Delta, {\rm cos}(2\phi) \rangle
(i\omega_{\nu},{\bf q}) &=& T \sum_n
\sum_{\bf p} {\rm Tr} [\delta\Delta \rho_3 G({\bf p},\omega_n)
{\rm cos}(2\phi) 
\nonumber\\
& & \;\; \times \rho_3 G({\bf p}-{\bf q},i\omega_n-i\omega_{\nu})] \ , \cr
\langle \delta\Delta, \delta\Delta \rangle
(i\omega_{\nu},{\bf q}) &=& T \sum_n
\sum_{\bf p} {\rm Tr} [\delta\Delta\rho_3 G({\bf p},\omega_n)
\delta\Delta
\nonumber\\
& & \;\; \times \rho_3 G({\bf p}-{\bf q},i\omega_n-i\omega_{\nu})] \ .
\label{corr}
\end{eqnarray}

After summing over the frequency,  the correlation functions are
written as
\begin{eqnarray}
Re \langle \delta \Delta,\gamma_{B1g} \rangle
&=&  N(0) \int_{\Delta}^{\infty} dE \frac{ \omega \Delta }{\sqrt{E^2-\Delta^2}}
\frac{1}{4 E^2-\omega^2}
\tanh{\frac{E}{2T}}
\nonumber\\
&=& \left[ N(0) \omega \Delta f \right],
\end{eqnarray}
where $f(\omega,T)$ is given by Eq. \ref{super}.
%
There is no imaginary part of the above correlation function.
On the other hand, the correlation function of the order parameter 
fluctuation is given by
\begin{eqnarray}
Re\langle \delta \Delta,\delta \Delta \rangle
&=& N(0) \int_{\Delta}^{\infty} \frac{dE}{\sqrt{E^2-\Delta^2}}
\frac{2E^2-\Delta^2}{\omega^2-4E^2}
\tanh{\frac{E}{2T}}
\nonumber\\
&=& g^{-1} -\left[ \left( \frac{\omega^2}{2}-\Delta^2 \right) f \right],
\end{eqnarray}
where
\begin{equation}
g^{-1} =
\frac{N(0)}{2} \int_{\Delta}^{\infty} \frac{dE}{\sqrt{E^2-\Delta^2}}
\tanh{\frac{E}{2T} }
\end{equation}
Now the imaginary part of the above correlation function is
\begin{eqnarray}
Im \langle \delta \Delta, \delta \Delta \rangle
&=& -  N(0) \frac{\pi}{2\omega}
\int_{\Delta}^{\infty} dE \frac{2E^2-\Delta^2}{\sqrt{E^2-\Delta^2}}
\tanh{\frac{E}{2T}}
\nonumber\\
& & \;\; \times
\left( \delta{(\omega- 2E)} + \delta{(\omega+2E)} \right)
\nonumber\\
&& \hspace{-1.5cm} =  N(0) \frac{\pi}{2\omega}
\frac{ \omega^2- 2\Delta^2}{\sqrt{\omega^2- 4 \Delta^2}}
\tanh{\frac{\omega}{4T}} \theta{(\omega- 2 \Delta)}.
\end{eqnarray}
Note that
$Im \langle \delta \Delta, \delta \Delta \rangle$ is $0$ for 
$\omega < 2\Delta$,
which is the regime we are interested in.
Therefore, the Raman intensity for $B_{1g}$ as well as $B_{2g}$ for
p-wave for $\omega < 2\Delta$ should be
\begin{eqnarray}
Im \left(  \frac{\langle [\gamma,\delta \Delta \rho_3] \rangle
\langle [\delta\Delta \rho_3,\gamma] \rangle}
{g^{-1} - \langle [\delta \Delta \rho_3,\delta \Delta \rho_3] \rangle }
\right)
&=& 
Im \frac{ 2 N(0) \omega^2 \Delta^2 f }
{ (\omega^2- 2\Delta^2) + i \Gamma}
\nonumber\\
&& \hspace{-2.0cm} =
2\pi N(0) \omega^2 \Delta^2 f  \delta{( \omega^2- 2\Delta^2)},
\end{eqnarray}
where a finite $\Gamma$ could come from the impurity scattering.
The clapping mode appears as a huge resonance with Lorentzian shape
(with a finite $\Gamma$) at the frequency of $\sqrt{2} \Delta$
shown in Fig. 2.

\subsection{the clapping mode and its contribution on Raman, $B_{1g,2g}$ in f-wave
superconductor}
Using the correlation functions, Eq. \ref{corr},
we obtained the following results for the f-wave superconductor.
\begin{eqnarray}
Re \langle \delta \Delta,\cos{(2\phi)} \rangle
&=&  \frac{2 N(0)\omega \Delta}{\pi} \int_{0}^{\infty} 
\frac{dE}{\omega^2-4E^2}
\nonumber\\
& &  \times 
\left( A (\frac{E}{\Delta}) \right) \tanh{\frac{E}{2T} }
\end{eqnarray}
Using the definition of $D(\omega,T)$, Eq. \ref{f_super}, the above
correlation function can be written as
\begin{equation}
Re \langle \delta \Delta,\cos{(2\phi)} \rangle
= N(0) \left[ \omega \Delta D(\omega, T) \right]
\end{equation}
On the other hand, the propagator of the clapping mode is given by
\begin{eqnarray}
Re \langle \delta \Delta,\delta \Delta \rangle
&=& g^{-1} -  \frac{2 N(0)}{\pi} \int_{0}^{\infty} 
\frac{dE}{4E^2- \omega^2}
\nonumber\\
& &  \times 
\left(  \frac{\omega^2}{2}  A (\frac{E}{\Delta}) 
-\Delta^2   B (\frac{E}{\Delta}) \right) 
\tanh{\frac{E}{2T} }
\nonumber\\
&=& g^{-1} - N(0) \frac{\omega^2-\omega_c^2}{2} D(\omega,T)
\nonumber\\
Im \langle \delta \Delta,\delta \Delta \rangle
&=&  \frac{N(0)}{\omega} 
\left( \omega^2  A (\frac{\omega}{2\Delta})
-2\Delta^2 B(\frac{\omega}{2\Delta})  \right)
\nonumber\\
& & \;\;\;\; \times
\tanh{\frac{\omega}{4T}},
\label{Im_f}
\end{eqnarray}
where
\begin{eqnarray}
A (k) &\equiv &
\frac{1}{\Delta} \left( K (k) - E(k) \right)
\nonumber\\
B(k) & \equiv  &
\frac{1}{3 \Delta} \left[ (k^2+2) K(k) -2 (k^2+1) E(k) \right] 
\nonumber\\
\omega_c & \equiv &  2 \Delta^2 B(k)/A(k).
\end{eqnarray}
Therefore, one can determine the position of the clapping mode, 
$\omega^R_c$,
using the following relations.
\begin{equation}
2 k^2 = \frac{1}{3} \left(
2 + \frac{k^2 K(k)-2 k^2 E(k)}{K(k)-E(k)} \right),
\end{equation}
where $k = \omega^R_c/(2\Delta)$.
This gives,
\begin{equation}
\omega^R_c = 1.02 \Delta.
\end{equation}

\end{document}